\documentclass[aps,prl,twocolumn,superscriptaddress,preprintnumbers,showpacs]{revtex4}
\usepackage{graphics}
\def\eq#1\en{\begin{equation} #1 \end{equation}}

\def\eqa#1\ena{\begin{eqnarray} #1 \end{eqnarray}}

\usepackage{graphicx,color}

\def\pp#1{\partial_{#1}}

\begin{document}

\preprint{BUTP-2003/02}
\preprint{CERN-TH/2003-012}
\preprint{IUB-TH/031}

\title{Neutrino dipole moments and charge radii in noncommutative space-time}
\author{P. Minkowski}
\affiliation{Theory Division, CERN, CH-1211 Geneva 23, Switzerland}
\affiliation{Institute for Theoretical Physics, University of Bern,
CH-3012 Bern, Switzerland}
\author{P. Schupp}
\affiliation{International University Bremen, 
Campus Ring 8,
28759 Bremen, Germany}
\author{J. Trampeti\'{c}}
\affiliation{Theory Division, CERN, CH-1211 Geneva 23, Switzerland}
\affiliation{Theoretical Physics Division, Rudjer Bo\v skovi\' c Institute, 
Zagreb, Croatia}
\affiliation{Theoretische Physik, Universit\"{a}t M\"{u}nchen, Theresienstr. 37, 80333 M\"{u}nchen, Germany}

\date{\today}

\begin{abstract}
In this paper we obtain a bound $\Lambda_{\rm NC}\stackrel{<}{\sim}\,150 $ TeV on the scale of space-time noncommutativity
considering photon-neutrino interactions. 
We compute ``$\star$-dipole moments'' and ``$\star$-charge radii'' originating from space-time noncommutativity 
and compare them with the dipole moments calculated in the neutrino-mass extended standard model (SM).
The computation
depends on the nature of the neutrinos, Dirac versus Majorana, their mass and the 
energy scale.  We focus on 
Majorana neutrinos. 
The ``$\star$--charge radius'' is found to be
$r^* = \sqrt{|\langle r^2_{\nu}\rangle_{\rm NC}|}=\left|3\sum_{i=1}^3 ({\theta}^{0i})^2\right|^{1/4}
\stackrel{<}{\sim} 1.6 \times 10^{-19}\; {\rm cm}$ 
at $\Lambda_{\rm NC} = 150$ TeV. 
\end{abstract}


\maketitle

In this paper we compare the consequences of the neutrino-photon interaction that can be induced by space-time 
non-commutativity~\cite{stwr}, with characteristic electromagnetic properties of neutrinos: charge radii
and dipole moments. These miniscule dipole moments are sensitive probes of fluctuations at scales as small as 
$10^{-35}\; {\rm cm}$, as seen through electromagnetic
interactions at long range.

The action of the model that we would like to study,
differs from commutative theory by the presence of $\star$-products
and Seiberg-Witten (SW) maps~\cite{SW,WESS,Zumino,calmet}. In the presence of space-time
noncommutativity,  neutral particles can couple to 
gauge bosons via a ${\star}$-commutator \cite{grosse} 
\begin{equation}
D^{\mbox{\scriptsize NC}}_\mu \widehat \psi = \partial_\mu \widehat \psi 
- i e \kappa \widehat A_\mu \star \widehat \psi
+ i e \kappa \widehat\psi \star \widehat A_\mu \;,
\label{1}
\end{equation}
where a hat denotes noncommutative fields that are  expanded in terms of regular fields
via SW maps.
The \mbox{$\star$-}products originate from antisymmetric tensor fields
that can conceivably be traced back to an extension of gravity.
In the language of quantized gravity, noncommutative effects belong to target space,
a quantum deformation of the classical base space.
On target space, the \mbox{$\star$-}products induce an algebraic structure of position 
operators that define noncommutative space-time.
In any case, observable effects are not necessarily fixed to the Planck scale.

The effective model of neutrino and photon interactions
in noncommutative space-time \cite{stwr}
provides a description of  the interaction of particles 
that enter from an asymptotically commutative
region into a noncommutative interaction region \cite{WESS}. 
The expansion in the form proposed 
in \cite{SW,WESS,Zumino,calmet} is understood as a perturbative description of
non-commutativity of the target space variables. 
Currently, this approach neither tries to 
describe dynamics of noncommutative structures
nor includes nonperturbative effects.
The action, written in terms of commutative fields, 
is gauge invariant under $U(1)_{\rm em}$-gauge transformations.
The requirements satisfied by our model are summarized in \cite{ws}.
For related work on noncommutative gauge theory and phenomenology,
see \cite{grosse,ws,HK,carl,tram}.

Expanding the $\star$-product in (\ref{1}) to first order in the antisymmetric
(Poisson) tensor $\theta^{\mu\nu}$, we find the following covariant derivative
on neutral spinor fields:
\begin{equation}
D^{\mbox{\scriptsize NC}}_\mu \widehat \psi = \partial_\mu \widehat \psi + 
e \kappa \theta^{\nu\rho} \, \partial_\nu\widehat A_\mu \, \partial_\rho \widehat \psi
\; .
\label{2}	
\end{equation}
We treat $\theta^{\mu\nu}$ as a constant background field of strength 
$|\theta^{\mu\nu}|=1/\Lambda^2_{\rm NC}$ that models the non-commutative structure
of space-time in the neighborhood of the interaction region \cite{carl}.
The scale of noncommutativity $\Lambda_{\rm NC}$ enters by choosing dimensionless matrix elements
$c^{\mu\nu}=\Lambda^2_{\rm NC} \theta^{\mu\nu}$ of order one.
As $\theta$ is not invariant under Lorentz transformations, the
neutrino field can pick up angular momentum in the interaction.
In the following we will assume $\theta$ to be constant in the mean over as 
large an interaction region as acceptable. The bounds that we are going to derive
rely on this assumption which we do not debate any further. How this regional
restriction can be derived is not of our concern here. However, if string theory leads 
consistently to associated Seiberg-Witten maps, it is for this theory to answer those
questions precisely.

The gauge-invariant action for a neutral fermion that couples to an Abelian gauge boson 
via (\ref{2}) is 
\begin{eqnarray}
S &= &\int d^4 x \,  \bar \psi 
\left[\left(\frac{}{}i\gamma^\mu \pp\mu  - m\right) 
\right.
\label{3}\\
&-&
\left.
\frac{e}{2}\kappa F_{\mu\nu}\left(
i \theta^{\mu\nu\rho}\pp\rho -\theta^{\mu\nu}m\right)\right]\psi\,,
\nonumber \\
{\theta}^{\mu\nu\rho} & = & {\theta}^{\mu\nu}\gamma^{\rho}+{\theta}^{\nu\rho}\gamma^{\mu}+
{\theta}^{\rho\mu}\gamma^{\nu}\,,
\nonumber
\end{eqnarray}
up to first order in $\theta$ \cite{stwr,tram}.
The noncommutative part of (\ref{3}) induces a {\it force}, proportional
to the gradient of the field strengths, which represents 
an interaction of the Stern-Gerlach type \cite{sg}. 
This interaction is non-zero even for $m_\nu=0$ and in this case reduces to the
coupling between the stress-energy tensor of the neutrino $T^{\mu\nu}$ and 
the symmetric tensor composed from $\theta$ and F \cite{stwr}.
Above interaction has also been derived in Ref. \cite{jabbari}, following the general discussion of 
the photon--electron noncommutative interaction in \cite{riad}. 
For a discussion of energy dependent Stern-Gerlach type of interactions,
we refer to~\cite{jabbari}. 

Following the general arguments of \cite{fy,Bernstein,shrock,mink} 
only the Dirac neutrino can have a magnetic moment. However,
the transition matrix elements relevant for $\nu_i \longrightarrow \nu_j$ may exist for both
Dirac and Majorana neutrinos.
In the neutrino-mass extended standard model \cite{mink}, the photon-neutrino effective vertex is
determined from the $\nu_i \longrightarrow \nu_j\,+\,\gamma$ transition, 
which is generated through 1-loop electroweak processes that
arise from the so-called ``neutrino--penguin'' diagrams via the exchange of $\ell=e,\mu,\tau$
leptons and weak bosons, and is given by \cite{fy,tram}
\begin{eqnarray}
J_{\mu}^{\rm eff}(\gamma\nu\bar\nu)\epsilon^{\mu}(q)&=&
\left\{ F_1(q^2) {\bar \nu_j}(p')(\gamma_{\mu}q^2-q_{\mu}{\not \!q})\nu_i(p)_L 
\right. \nonumber\\
&-&\left.iF_2(q^2)\left[ m_{\nu_j}{\bar \nu_j}(p')\sigma_{\mu\nu}q^{\nu}\nu_i(p)_L 
\right.\right. \nonumber\\
&+&\left.\left. m_{\nu_i}{\bar \nu_j}(p')\sigma_{\mu\nu}q^{\nu}\nu_i(p)_R\right]\right\}
\epsilon^{\mu}(q).
\label{4}
\end{eqnarray}
The above effective interaction is invariant under electromagnetic gauge transformations.
The first term in (\ref{4}) vanishes identically for real photon due to the electromagnetic gauge condition. 

From the general decomposition of the second term of the transition matrix element T obtained from (\ref{4}),
\begin{eqnarray}
{\rm T}=-i\epsilon^{\mu}(q){\bar \nu}(p')\left[A(q^2)-B(q^2)\gamma_5\right]\sigma_{\mu\nu}q^{\nu}\nu(p),
\label{5}
\end{eqnarray}
we found the following expression for the electric and magnetic dipole moments 
\begin{eqnarray}
d^{\rm el}_{ji}\equiv B(0)\hspace{-2mm}&=&\hspace{-2mm}\frac{-e}{M^{*2}}\left( m_{\nu_i}-m_{\nu_j}\right)
\hspace{-3mm}\sum_{\ell=e,\mu,\tau}\hspace{-3mm}{\rm U}^{\dagger}_{jk}{\rm U}^{}_{ki}
F_2(\frac{m^2_{\ell_k}}{m^2_W}),
\label{6} \\
\mu_{ji}\equiv A(0)\hspace{-2mm}&=&\hspace{-2mm}\frac{-e}{M^{*2}}\left( m_{\nu_i}+m_{\nu_j}\right)
\hspace{-3mm}\sum_{\ell=e,\mu,\tau}\hspace{-3mm}{\rm U}^{\dagger}_{jk}{\rm U}^{}_{ki}
F_2(\frac{m^2_{\ell_k}}{m^2_W}),
\label{7}
\end{eqnarray}
where $i,j,k=1,2,3$ denotes neutrino species, and  
\begin{eqnarray}
F_2(\frac{m^2_{\ell_k}}{m^2_W})\simeq -\frac{3}{2}+\frac{3}{4}\frac{m^2_{\ell_k}}{m^2_W},
\;\;\;\frac{m^2_{\ell_k}}{m^2_W}\ll 1,
\label{8}
\end{eqnarray}
was obtained after the loop integration. In Eqs. (\ref{6}) and (\ref{7}) 
$M^*=4\pi\,v=3.1$ TeV, where $v=(\sqrt 2\, G_F)^{-1/2}=246$ GeV 
represents the vacuum expectation value of the scalar Higgs field \cite{DMT}. 

The neutrino mixing matrix U \cite{MNS} is governing the decomposition of a coherently
produced left-handed neutrino $\widetilde{\nu}_{L,\ell}$ 
associated with charged-lepton-flavor $\ell = e, \mu, \tau$ into
the mass eigenstates $\nu_{L,i}$:
\begin{eqnarray}
|\widetilde{\nu}_{L,\ell};\,\vec p\,\rangle =
\sum_i {\rm U}_{\ell i} |\nu_{L,i};\,\vec p ,m_i\,\rangle ,
\label{9}
\end{eqnarray}
For a Dirac neutrino $i=j$ \cite{Bernstein,FS}. Using $m_{\nu} =0.05\,{\rm eV}$ \cite{nobel}, 
and with the definition $\mu_{ii}\equiv \mu_{\nu_i}$, from (\ref{7}),
in units of [$\rm e\,cm$] and Bohr magneton, we obtain 
\begin{eqnarray}
\mu_{\nu_i}&=&\frac{3e}{M^{*2}}m_{\nu_i}\left[1-\frac{1}{2}
\sum_{\ell=e,\mu,\tau} \frac{m^2_\ell}{m^2_W}\, |{\rm U}_{\ell i}|^2\right]
\nonumber \\
{}&\simeq&  1.56\times 10^{-26} [\rm e/eV]=0.29 \times 10^{-30} \,[\rm e\;cm] 
\nonumber\\
{}&=& 1.60\times 10^{-20} \mu_B.
\label{10}
\end{eqnarray}
From formula (\ref{10}) it is clear that the
chirality flip, which is necessary to induce the magnetic moment, arises only from the neutrino masses.
The Dirac neutrino magnetic moment (\ref{10}) is
still much smaller than the bounds obtained from astrophysics \cite{FY,AOT}.
More details about Dirac neutrinos can be found in \cite{VV,VVO}.

In the case of off-diagonal transition moments, the first term in (\ref{8}) 
vanishes in the summation over $\ell$ due to the orthogonality condition of U (GIM cancellation)
\begin{eqnarray}
\hspace{-2mm}d^{\rm el}_{{\bar\nu_j}{\nu_i}}\hspace{-1.5mm}&=&\hspace{-1.5mm}\frac{3e}{2M^{*2}} 
\left( m_{\nu_i}-m_{\nu_j}\right)
\sum_{\ell=e,\mu,\tau}\frac{m^2_{\ell_k}}{m^2_W}
{\rm U}^{\dagger}_{jk}{\rm U}^{}_{ki},
\label{12}\\
\hspace{-2mm}{\mu}_{{\bar\nu_j}{\nu_i}}\hspace{-1.5mm}&=&\hspace{-1.5mm}\frac{3e}{2M^{*2}}
\left( m_{\nu_i}+m_{\nu_j}\right) 
\sum_{\ell=e,\mu,\tau}\frac{m^2_{\ell_k}}{m^2_W} {\rm U}^{\dagger}_{jk}{\rm U}^{}_{ki} .
\label{13}
\end{eqnarray}
In Majorana 4-component notation
the Hermitian, neutrino-flavor antisymmetric, electric and magnetic dipole operators are
\begin{eqnarray}
{D_5 \choose D}^{\mu\nu}_{ij} = 
e\,\psi^{\top}_{\rm i}\left[\rm C \;\sigma^{\mu\nu}{\gamma_5 \choose i \mbox{\bf 1}}\right]\,\psi_{\rm j}\;.
\label{14}
\end{eqnarray}
The characterizing feature of Majorana neutrinos, i.e.\ fields that do not distinguish particle 
from antiparticle ($\psi_i=\psi_i^c$), forces one 
to use both charged lepton and antilepton propagators in the loop calculation of 
neutrino-penguin diagrams, producing
a transition matrix element T which is a complex antisymmetric quantity in lepton-flavor space:
\begin{eqnarray}
{\rm T_{ji}}&=&-i\epsilon^{\mu}{\bar\nu_j}
\left[(A_{ji}-A_{ij})-(B_{ji}-B_{ij})\gamma_5\right]\sigma_{\mu\nu}q^{\nu}\nu_i
\nonumber\\
&=&-i\epsilon^{\mu}{\bar\nu_j}
\left[2{\rm iIm}A_{ji}-2{\rm Re}B_{ji}\gamma_5\right]\sigma_{\mu\nu}q^{\nu}\nu_i.
\label{15}
\end{eqnarray}
From this equation it is explicitly clear that for $i=j$, $d^{\rm el}_{\nu_i}={\mu}_{\nu_i}=0$.
Also, considering transition moments, only one of two terms in (\ref{15}) is 
non-vanishing if the interaction respects CP invariance: The first term vanishes if 
the relative CP of $\nu_i$ and $\nu_j$ is even, and the second term vanishes if it is odd \cite{shrock}.
Finally, the dipole moments describing the transition from Majorana neutrino mass eigenstate-flavor 
$\nu_j$ to $\nu_k$ in the mass extended standard model are:
\begin{eqnarray}
\hspace{-2mm}d^{\rm el}_{{\nu_i}{\nu_j}}\hspace{-1.5mm}&=&\hspace{-1.5mm}\frac{3e}{2M^{*2}} 
\left( m_{\nu_i}-m_{\nu_j}\right)
\hspace{-2mm}\sum_{\ell=e,\mu,\tau}\frac{m^2_{\ell_k}}{m^2_W}
{\rm Re}{\rm U}^{\dagger}_{jk}{\rm U}^{}_{ki},
\label{16}\\
\hspace{-2mm}{\mu}_{{\nu_i}{\nu_j}}\hspace{-1.5mm}&=&\hspace{-1.5mm}\frac{3e}{2M^{*2}} 
\left( m_{\nu_i}+m_{\nu_j}\right)
\hspace{-2mm}\sum_{\ell=e,\mu,\tau}\frac{m^2_{\ell_k}}{m^2_W} 
{\rm i\,Im}{\rm U}^{\dagger}_{jk}{\rm U}^{}_{ki} ,
\label{17}
\end{eqnarray}
For the Majorana case the neutrino-flavor mixing matrix U is approximatively unitary, i.e  
it is necessarily of the following form \cite{DMT}
\begin{eqnarray}
\sum_{i=1}^3 {\rm U}^{\dagger}_{jk}{\rm U}^{}_{ki} = {\delta}_{ji} - \varepsilon_{ji},
\label{18}
\end{eqnarray}
where $\varepsilon$ is a hermitian nonnegative matrix (i.e. with all eigenvalues nonnegative) and 
\begin{eqnarray}
|\varepsilon|= \sqrt{{\rm Tr}\;\varepsilon^2} &=& {\cal O} \; (m_{\nu_{\rm light}}/m_{\nu_{\rm heavy}}),
\nonumber \\
&\sim& 10^{-22} \;\, {\rm to}\;\, 10^{-21}. 
\label{19}
\end{eqnarray}
The case $|\varepsilon|=0$ is excluded by the very existence of oscillation effects.

The neutrino dipole moments (\ref{16},\ref{17}) violate lepton number by $\pm2$ and for 
a general neutrino mass matrix, they independently violate CP.
Finally, note that there is no exact GIM cancellation in lepton-flavor space, 
at least in the Majorana case considered here, unlike the case for off--diagonal transition matrix elements
(\ref{12},\ref{13}), and for quark-flavors \cite{tram}.

The flavor cancellation mechanism operates partly for the Majorana light neutrino dipole moments. 
It is generated through the smallness of light neutrino masses controlled by their heavy counterparts,
combined with the quadratic charged lepton mass asymmetry in (\ref{8},\ref{19}). 

Note that expressions (\ref{4}) to (\ref{19}) are exact 
in their bilinear dependence on the mixing matrix U$_{kj}$ to 
all orders of light and inverse heavy neutrino masses.
However, the mixing matrix U is subunitary, as it must be a submatrix of the truly unitary $6\times 6$
neutrino flavor mixing matrix.

The electromagnetic dipole moments universally reduce to the three light mass 
eigenfields (eigenstates) of neutrinos and antineutrinos,
even though the mediating interactions proceed through light and heavy
weak-interaction eigenfields (eigenstates) involving
minimally six flavors.
This extends the analogous situation pertaining to neutrino and antineutrino 
oscillations, valid at energies, in production and 
detection, far below the masses of heavy--neutrino flavors \cite{mink}. 
The MNS parameterization is generated by diagonalizing the light--flavor 
mass matrix in (\ref{6},\ref{7}). 
The corresponding analytic structure is quite definite,
yet often globally referred to as the see--saw mechanism \cite{gell}.
Specific mass patterns for light-neutrino flavors arising from approximate discrete
symmetries are discussed in \cite{fri}.

The transition dipole moments in general receive very small contributions 
because of the smallness of the neutrino mass, 
$|m_{\nu}|\simeq 10^{-2}$ eV \cite{nobel}. The largest contribution amog them is proportional to Re and Im part
of ${\rm U}^{\dagger}_{3\tau}{\rm U}^{}_{\tau 2}$, which corresponds to the $2 \to 3$ transition.

For the sum and difference of neutrino masses we assume hierarchical structure 
and take $|m_3 + m_2| \simeq |m_3 - m_2| \simeq |\Delta m_{32}^2|^{1/2} = 0.05$ eV \cite{nobel}. 
For the MNS matrix elements we set $|{\rm Re}{\rm U}^{\dagger}_{3\tau}{\rm U}^{}_{\tau 2}| 
\simeq|{\rm Im}{\rm U}^{\dagger}_{3\tau}{\rm U}^{}_{\tau 2}| \le 0.5$.
The electric and magnetic transition dipole moments of neutrinos 
$d^{\rm el}_{\nu_2\nu_3}$ and $\mu_{\nu_2\nu_3}$ are then denoted as $\left(d^{\rm el}_{\rm mag}\right)_{23}$  
and are given by
\begin{eqnarray}
\left|\left(d^{\rm el}_{\rm mag}\right)_{23}\right| &=& \;\frac{3e}{2M^{*2}}\;
\frac{m^2_{\tau}}{m^2_W}\sqrt{|\Delta m_{32}^2|}
{|{\rm Re}{\rm U}^{\dagger}_{3\tau}{\rm U}^{}_{\tau 2}|\choose |{\rm Im}{\rm U}^{\dagger}_{3\tau}{\rm U}^{}_{\tau 2}|},
\nonumber \\
&\stackrel{<}{\sim}& 2.03 \times 10^{-30} [\rm e/eV] = 0.38 \times 10^{-34} \,[\rm e\;cm],
\nonumber \\
&=& 2.07\times 10^{-24}\, \mu_B.
\label{20}
\end{eqnarray}
The electric transition dipole moments of light neutrinos are smaller than the ones of the d-quark. 
This is {\it the} order of magnitude of light neutrino transition dipole moments underlying the see--saw mechanism. 
It is by orders of magnitude smaller than in unprotected SUSY models.

Now we extract an upper limit on the $\star$-gradient interaction. 
The strength of the interaction (\ref{3}) becomes $|m_{\nu}\,e\,\kappa\,\theta \,F|$. 
We compare it with the dipole
transition interactions $|F\, d^{\rm el}_{\rm mag}|$ for Majorana case (\ref{16},\ref{17}).
Assuming that contributions from the 
neutrino-mass extended standard model are at least as large as those from noncommutativity, for $\kappa = 1$ we
derive the following bound on noncommutativity 
arising from the Majorana nature of neutrinos:
\begin{eqnarray}
\Lambda_{\rm NC} &\stackrel{>}{\sim}& 
\left|{\frac{e\,\kappa\,m_{\nu}}{\left(d^{\rm el}_{\rm mag}\right)_{23}}}\right|^{1/2}\,
\simeq 150 \,{\rm TeV}.
\label{22}
\end{eqnarray}
This is the main result of our considerations \cite{foot1}. 
In Ref. \cite{jabbari} the neutrino energy dependence was taken into account.
The bound on noncommutativity thus obtained is not a strict lower limit, 
but rather indicate the scale $\Lambda_{\rm NC}$
at which the expected values of the neutrino electromagnetic dipole moments 
due to noncommutativity in our model matches the standard model contributions.
We would like to point out that on the scale of noncommutativity this bound involves
only the basic properties of neutrinos and photons. 

We proceed with determination of the radius of the photon--neutrino interaction, 
by evaluating the quantity that we shall call 
the neutrino $\star$-charge radius $r^{*2} = \langle r^2_{\nu}\rangle_{\rm NC}$.  
Since noncommutativity can be a source of ``transvers plasmon'' decay 
into neutrino--antineutrino pairs \cite{stwr}, this is to be compared with 
the same process induced by the neutrino charge radii defined by the axial
electromagnetic interaction form factor \cite{shrock,Bernstein,Altherr,dsm,as,salati,BPV,fs} 
in the neutrino-mass extended standard model:
\begin{eqnarray}
\langle r_{\nu}^2\rangle =6\left[{\frac {\partial F_1(q^2)}{\partial q^2}}\right]_{q^2=0}\,;
\left[F_1(q^2)\right]_{q^2 \to 0} \longrightarrow
\frac{q^2}{6}\langle r_{\nu}^2\rangle,
\label{23}
\end{eqnarray}
which in the limit of massless neutrinos corresponds to
\begin{eqnarray}
\langle r^2_{\nu_{\ell}}\rangle \cong \hspace*{-1mm}\frac{2}{M^{*2}} \hspace*{-1mm}\left(3-2{\rm log} 
\frac{m^2_{\ell}}{m^2_W}\hspace*{-1mm}\right)\hspace*{-1mm}
=\hspace*{-1mm}\frac{G_F}{\sqrt2\,\pi^2} \hspace*{-1mm}
\left(\frac{3}{4}+{\rm log} \frac{m_W}{m_{\ell}}\hspace*{-1mm}\right)
\hspace*{-1mm}.
\label{24}
\end{eqnarray}
We estimate the charge radii in the standard model from (\ref{24}) by taking $\ell=e$:
$\sqrt{|\langle r^2_{\nu_e}\rangle|} \simeq 0.64\times \,10^{-16}$ [cm]. Here we remark that astrophysical
estimates give interesting bounds \cite{gm1,gm,hnr,okada}.
These calculations should implement all neutrino flavor properties. The so derived bounds may also
help in establishing the Majorana nature of light neutrinos.

To estimate the $\star$-charge radii we first evaluate the partial width
\begin{eqnarray}
\sum_{\ell=e,\mu,\tau}\Gamma_{\rm SM}(\gamma \rightarrow {\bar\nu}_{\ell}^{\rm L}\nu_{\ell}^{\rm L})=
\frac{\alpha}{144}\frac{q^6}{E_{\gamma}}\sum_{\ell=e,\mu,\tau}\left|\langle r^2_{\nu_{\ell}}\rangle\right|^2,
\label{25}
\end{eqnarray}
which gives the SM rate induced by the charge radii \cite{as}. 

For plasmon at rest $q^2=E^2_{\gamma}=\omega^2_{\rm pl}$. 
Taking the average value of the plasmon frequencies of red-giant and
white-dwarf stars $\omega_{\rm pl}=15$ keV \cite{as,salati}, we obtain 
\begin{eqnarray}
\Gamma^{-1}_{\rm SM}(\gamma \rightarrow \bar\nu \nu)&=&
\left(\frac{1\,{\rm keV}}{\omega_{\rm pl}}\right)^5 
\times 0.25 \times10^{13}\; {\rm years}
\nonumber \\
&\simeq& 3\times 10^6 \; {\rm years}.
\label{25a}
\end{eqnarray} 
This value has to be compared with astrophysical observations.

The next step is to compare (\ref{25}) with the noncommutative rate
$\sum\Gamma_{\rm NC}(\gamma \rightarrow {\bar{\nu}}_{\ell}^{\rm L} \nu_{\ell}^{\rm L})$.

From Eq. (\ref{3}) we extract  
the following gauge-invariant amplitude for  
the ${\gamma}(q)\to {\nu}(k'){\bar {\nu}}(k)$ vertex in momentum space for the left--chiral neutrinos:
\begin{eqnarray}
{\cal M}_{\gamma\nu\bar\nu}  =  ie\,\kappa\,\bar\psi_L({\theta}^{\mu\nu\rho}k_{\nu}q_{\rho})\psi_L\,\epsilon_{\mu}(q).
 \label{26}
\end{eqnarray}
The amplitude (\ref{25a})
for the off-shell photon decay to massless Majorana neutrinos, 
leads to the following rate in the c.m. system \cite{stwr}:
\begin{eqnarray}
&&\sum_{\ell=e,\mu,\tau}\Gamma_{\rm NC}(\gamma\rightarrow {\bar{\nu}}_{\ell}^{\rm L} \nu_{\ell}^{\rm L})
= \frac{\alpha}{16}\frac{\kappa^2\,q^6}{{\rm E}_{\gamma}\Lambda^4_{\rm NC}}\sum_{i=1}^{3}(c^{0i})^2 .
\label{27}
\end{eqnarray}
The coefficients, ($c^{0i}$), are not independent. In pulling out the overall scale 
$\Lambda_{\rm NC}$, we can always
impose the constraint $ \sum_{i=1}^3 (c^{0i})^2 \equiv {\vec E}^2_{\theta}=1$ \cite{HK}. 

We obtain the $\star$--charge radii, which in fact could
represent the range of noncommutativity, to be a simple function of the scale of noncommutativity:
\begin{equation}
r^* = \sqrt{|\langle r^2_{\nu}\rangle_{\rm NC}|} = \frac{\sqrt{\sqrt{3}\,\kappa}}{\Lambda_{\rm NC}}.
\label{28}
\end{equation}
The bound from Majorana neutrino induced scale of noncommutativity (\ref{22}),
for $\kappa=1$ implies \cite{foot2}
\begin{eqnarray} 
r^* \stackrel{<}{\sim} 1.6 \times 10^{-19} [{\rm cm}].
\label{29}
\end{eqnarray}
This  means that the $\star$--induced charge radii $r^*$
at the $\Lambda_{\rm NC}\,\stackrel{>}{\sim} 150 \,{\rm TeV}$ scale,
are dominated by the neutrino-mass extended standard model physics and are practically unobservable. 

Note that there are polarization phenomena induced by the noncommutativity
tensor $\theta^{\mu\nu}$, which would involve correlations between spin and momenta. 
These, however, have been integrated out in our estimate. The motivation to do so lies in the fact that 
our model for the photon--neutrino interaction represents only the tree-level effective 
noncommutative gauge field theory in which the question of renormalization 
is not addressed \cite{stwr,jabbari,riad,UV/IR,luis}. 

In conclusion, we have compared the neutrino-mass extended standard model 
charge radii and electromagnetic dipole moments
of neutrinos with their analogs arising from a theory of noncommutative space-time.
If the charge radii and electromagnetic dipole moments 
should be found experimentally different from those predicted by 
the neutrino-mass extended standard model, as indicated from astrophysics \cite{gm1,gm,hnr,okada}, then this 
could be a signature of noncommutativity. 
We observe that the sensitivity to noncommutativity in our model involving neutrinos \cite{stwr} 
appears to be a function of the scale of energy involved in the physical process,
running from the weak scale \cite{stwr} up to a few hundreds of TeV's [this work]. 
In this way we can ``understand'' neutrinos as particles which manifest them self as Majorana objects at 
the short distances (high energies).
 
Independently of the noncommutative part of the story,
we believe that the difference between (\ref{10}) and (\ref{20}),
produced by standard model physics, points toward the right direction for the determination of 
the real nature of neutrinos.  
We hope that this comparison sheds light on the magnitude of associated phenomena thus induced.
Our results relate to a set of physics problems that involve mass and electromagnetic
properties of neutrinos. 

\vspace{.5cm}

We would like to thank  G. Raffelt for suggesting the study of the neutrino charge radii
and J. Wess for many helpful discussions leading to the construction of our model. 

The work of P.M. is supported by the Swiss National Science Foundation.
The work of J.T. is supported by the Ministry of Science and Technology 
of Croatia under Contract No. 0098002.


\begin{thebibliography}{99}
\bibitem{foot1}
                The neutrino mass extended standard model, as a consequence of loops (\ref{8}),
                produces four orders of magnitude different
                dipole moments for a Dirac neutrino (\ref{10}) and a Majorana neutrino (\ref{20}),
                leading also to two different bounds on the scale of noncommutativity. 
                The difference in complex structure between Dirac and Majorana neutrinos
		leads in SM loop calculations to an 
                $(m^2_{\ell}/m^2_W)$-suppression of Majorana dipole moments 
                relative to the Dirac ones. 
                On the other hand, 
                our model (\ref{3}) does not distinguish at tree-level between Dirac and Majorana 
                neutrinos, so that there is no compensation of the differences coming 
		from the SM computation.
\bibitem{foot2}	In the case of Dirac neutrinos, the bound on the non-commutativity scale is
                $\Lambda_{\rm NC}\,\stackrel{>}{\sim} 1.8 \,{\rm TeV}$ and the $\star$--induced charge radius
		is $r^* \hspace{-1mm}\stackrel{<}{\sim} 0.14 \times 10^{-16} [{\rm cm}]$, which
		is about factor of five below the SM value (\ref{24}).
\end{thebibliography}
\end{document}